\documentclass[aps,prl,superscriptaddress,twocolumn,showpacs,floatfix]{revtex4}

\usepackage{amsmath}
\usepackage{graphicx}
\usepackage{dcolumn}
\usepackage{bm}
\usepackage{array}
\newcolumntype{M}[1]{>{\centering\arraybackslash}m{#1}}
\newcolumntype{N}{@{}m{0pt}@{}}

\def\lsim{\mathrel{\raise.3ex\hbox{$<$\kern-.75em\lower1ex\hbox{$qf$}}}}
\def\gsim{\mathrel{\raise.3ex\hbox{$>$\kern-.75em\lower1ex\hbox{$\sim$}}}}

\begin{document}


\title{Dark Matter Search Results from the PICO-2L C$_3$F$_8$ Bubble Chamber}

\author{C.~Amole}
\affiliation{Department of Physics, Queen's University, Kingston, Ontario K7L 3N6, Canada}
\author{M.~Ardid}
\affiliation{Universitat Polit\`ecnica de Val\`encia, IGIC, 46730 Gandia, Spain}
\author{D.~M.~Asner}
\affiliation{Pacific Northwest National Laboratory, Richland, Washington 99354, USA}
\author{D.~Baxter}
\affiliation{Department of Physics and Astronomy, Northwestern University, Evanston, Illinois 60208, USA}
\author{E.~Behnke}
\affiliation{Department of Physics and Astronomy, Indiana University South Bend, South Bend, Indiana 46634, USA}
\author{P.~Bhattacharjee}
\affiliation{Saha Institute of Nuclear Physics, Astroparticle Physics and Cosmology Division, Kolkata, West Bengal 700064, India}
\author{H.~Borsodi}
\affiliation{Department of Physics and Astronomy, Indiana University South Bend, South Bend, Indiana 46634, USA}
\author{M.~Bou-Cabo}
\affiliation{Universitat Polit\`ecnica de Val\`encia, IGIC, 46730 Gandia, Spain}
\author{S.~J.~Brice}
\affiliation{Fermi National Accelerator Laboratory, Batavia, Illinois 60510, USA}
\author{D.~Broemmelsiek}
\affiliation{Fermi National Accelerator Laboratory, Batavia, Illinois 60510, USA}
\author{K.~Clark}
\affiliation{Department of Physics, University of Toronto, Toronto, Ontario  M5S 1A7, Canada}

\author{J.~I.~Collar}
\affiliation{Enrico Fermi Institute, KICP and Department of Physics,
University of Chicago, Chicago, Illinois 60637, USA}
\author{P.~S.~Cooper}
\affiliation{Fermi National Accelerator Laboratory, Batavia, Illinois 60510, USA}
\author{M.~Crisler}
\affiliation{Fermi National Accelerator Laboratory, Batavia, Illinois 60510, USA}
\author{C.~E.~Dahl}
\affiliation{Department of Physics and Astronomy, Northwestern University, Evanston, Illinois 60208, USA}
\affiliation{Fermi National Accelerator Laboratory, Batavia, Illinois 60510, USA}
\author{S.~Daley}
\affiliation{Department of Physics, Queen's University, Kingston, Ontario K7L 3N6, Canada}
\author{M.~Das}
\affiliation{Saha Institute of Nuclear Physics, Astroparticle Physics and Cosmology Division, Kolkata, West Bengal 700064, India}
\author{F.~Debris}
\affiliation{D\'epartement de Physique, Universit\'e de Montr\'eal, Montr\'eal, Qu\'ebec H3C 3J7, Canada}
\author{N.~Dhungana}
\affiliation{Department of Physics, Laurentian University, Sudbury, Ontario P3E 2C6, Canada}
\author{J.~Farine}
\affiliation{Department of Physics, Laurentian University, Sudbury, Ontario P3E 2C6, Canada}
\author{I.~Felis}
\affiliation{Universitat Polit\`ecnica de Val\`encia, IGIC, 46730 Gandia, Spain}
\author{R.~Filgas}
\affiliation{Institute of Experimental and Applied Physics, Czech Technical University in Prague, Prague, 12800, Czech Republic}
\author{M.~Fines-Neuschild}
\affiliation{D\'epartement de Physique, Universit\'e de Montr\'eal, Montr\'eal, Qu\'ebec H3C 3J7, Canada}
\author{F.~Girard}
\affiliation{D\'epartement de Physique, Universit\'e de Montr\'eal, Montr\'eal, Qu\'ebec H3C 3J7, Canada}
\author{G.~Giroux}
\affiliation{Department of Physics, Queen's University, Kingston, Ontario K7L 3N6, Canada}
\author{M.~Hai}
\affiliation{Enrico Fermi Institute, KICP and Department of Physics,
University of Chicago, Chicago, Illinois 60637, USA}
\author{J.~Hall}
\affiliation{Pacific Northwest National Laboratory, Richland, Washington 99354, USA}
\author{O.~Harris}
\affiliation{Department of Physics and Astronomy, Indiana University South Bend, South Bend, Indiana 46634, USA}
\author{C.~M.~Jackson}
\affiliation{D\'epartement de Physique, Universit\'e de Montr\'eal, Montr\'eal, Qu\'ebec H3C 3J7, Canada}
\author{M.~Jin}
\affiliation{Department of Physics and Astronomy, Northwestern University, Evanston, Illinois 60208, USA}
\author{C.~B.~Krauss}
\affiliation{Department of Physics, University of Alberta, Edmonton, Alberta T6G 2G7, Canada}
\author{M.~Lafreni\`ere}
\affiliation{D\'epartement de Physique, Universit\'e de Montr\'eal, Montr\'eal, Qu\'ebec H3C 3J7, Canada}
\author{M.~Laurin}
\affiliation{D\'epartement de Physique, Universit\'e de Montr\'eal, Montr\'eal, Qu\'ebec H3C 3J7, Canada}
\author{I.~Lawson}
\affiliation{SNOLAB, Lively, Ontario, P3Y 1N2, Canada}
\affiliation{Department of Physics, Laurentian University, Sudbury, Ontario P3E 2C6, Canada}
\author{I.~Levine}
\affiliation{Department of Physics and Astronomy, Indiana University South Bend, South Bend, Indiana 46634, USA}
\author{W.~H.~Lippincott}
\affiliation{Fermi National Accelerator Laboratory, Batavia, Illinois 60510, USA}
\author{E.~Mann}
\affiliation{Department of Physics and Astronomy, Indiana University South Bend, South Bend, Indiana 46634, USA}
\author{J.~P.~Martin}
\affiliation{D\'epartement de Physique, Universit\'e de Montr\'eal, Montr\'eal, Qu\'ebec H3C 3J7, Canada}
\author{D.~Maurya}
\affiliation{Bio-Inspired Materials and Devices Laboratory (BMDL), Center for Energy Harvesting Materials and Systems (CEHMS), Virginia Tech, Blacksburg, Virginia 24061, USA}

\author{P.~Mitra}
\affiliation{Department of Physics, University of Alberta, Edmonton, Alberta T6G 2G7, Canada}
\author{R.~Neilson}
\affiliation{Enrico Fermi Institute, KICP and Department of Physics,
University of Chicago, Chicago, Illinois 60637, USA}
\affiliation{Department of Physics, Drexel University, Philadelphia, Pennsylvania 19104, USA}
\author{A.~J.~Noble}
\affiliation{Department of Physics, Queen's University, Kingston, Ontario K7L 3N6, Canada}
\author{A.~Plante}
\affiliation{D\'epartement de Physique, Universit\'e de Montr\'eal, Montr\'eal, Qu\'ebec H3C 3J7, Canada}
\author{R.~B.~Podviianiuk}
\affiliation{Department of Physics, Laurentian University, Sudbury, Ontario P3E 2C6, Canada}
\author{S.~Priya}
\affiliation{Bio-Inspired Materials and Devices Laboratory (BMDL), Center for Energy Harvesting Materials and Systems (CEHMS), Virginia Tech, Blacksburg, Virginia 24061, USA}
\author{A.~E.~Robinson}
\affiliation{Enrico Fermi Institute, KICP and Department of Physics,
University of Chicago, Chicago, Illinois 60637, USA}
\author{M.~Ruschman}
\affiliation{Fermi National Accelerator Laboratory, Batavia, Illinois 60510, USA}
\author{O.~Scallon}
\affiliation{Department of Physics, Laurentian University, Sudbury, Ontario P3E 2C6, Canada}
\affiliation{D\'epartement de Physique, Universit\'e de Montr\'eal, Montr\'eal, Qu\'ebec H3C 3J7, Canada}
\author{S.~Seth}
\affiliation{Saha Institute of Nuclear Physics, Astroparticle Physics and Cosmology Division, Kolkata, West Bengal 700064, India}
\author{A.~Sonnenschein}
\affiliation{Fermi National Accelerator Laboratory, Batavia, Illinois 60510, USA}
\author{N.~Starinski}
\affiliation{D\'epartement de Physique, Universit\'e de Montr\'eal, Montr\'eal, Qu\'ebec H3C 3J7, Canada}
\author{I.~\v{S}tekl}
\affiliation{Institute of Experimental and Applied Physics, Czech Technical University in Prague, Prague, 12800, Czech Republic}
\author{E.~V\'azquez-J\'auregui}
\affiliation{Instituto de F\'isica, Universidad Nacional Aut\'onoma de M\'exico, M\'exico D. F. 01000, M\'exico}
\affiliation{SNOLAB, Lively, Ontario, P3Y 1N2, Canada}
\affiliation{Department of Physics, Laurentian University, Sudbury, Ontario P3E 2C6, Canada}
\author{J.~Wells}
\affiliation{Department of Physics and Astronomy, Indiana University South Bend, South Bend, Indiana 46634, USA}
\author{U.~Wichoski}
\affiliation{Department of Physics, Laurentian University, Sudbury, Ontario P3E 2C6, Canada}
\author{V.~Zacek}
\affiliation{D\'epartement de Physique, Universit\'e de Montr\'eal, Montr\'eal, Qu\'ebec H3C 3J7, Canada}
\author{J.~Zhang}
\affiliation{Department of Physics and Astronomy, Northwestern University, Evanston, Illinois 60208, USA}

\collaboration{PICO Collaboration}
\noaffiliation

\date{\today}


\begin{abstract}
New data are reported from the operation of a 2-liter C$_3$F$_{8}$ bubble
chamber in the SNOLAB underground laboratory, with a total exposure of 211.5 kg-days at four different energy thresholds below 10 keV. These data show that C$_3$F$_8$ provides excellent electron-recoil and alpha rejection capabilities at very low thresholds. The chamber exhibits an electron-recoil sensitivity of $<3.5\times10^{-10}$ and an alpha rejection factor of $>98.2\%$. These data also include the first observation of a dependence of acoustic signal on alpha energy. Twelve single nuclear recoil event candidates were observed during the run. The candidate events exhibit timing characteristics that are not consistent with the hypothesis of a uniform time distribution, and no evidence for a dark matter signal is claimed.  These data provide the most sensitive direct detection constraints on WIMP-proton spin-dependent 
scattering to date, with significant sensitivity at low WIMP masses for spin-independent WIMP-nucleon scattering.
\end{abstract}

\pacs{29.40.-n, 95.35.+d, 95.30.Cq,   FERMILAB-PUB-14-456-AE-E}
\maketitle

Understanding the nature of dark matter is one of the most important goals in modern particle physics~\cite{P5,Snowmass}. A leading candidate to explain the dark matter is a relic density of cold, nonbaryonic weakly interacting massive particles or WIMPs, and direct detection dark matter experiments hope to observe the nuclei recoiling from the rare collisions of WIMPs with ordinary matter~\cite{dmevidence,Jungman,wimptheory,wimpdetection}. Historically, the interaction of dark matter with normal matter has been divided into two categories, spin dependent (SD) and spin independent (SI).

The superheated detector technology has been at the forefront of SD searches~\cite{PRD,previousPRL,PICASSOlimit,simple2014}, using  refrigerant targets including CF$_3$I, C$_4$F$_{10}$ and C$_2$ClF$_5$, and two primary types of detectors: bubble chambers and droplet detectors. The PICO Collaboration (formed from the merger of PICASSO and COUPP) has now operated a 2.90~kg C$_3$F$_8$ bubble chamber from October 2013 to May 2014 in the SNOLAB underground laboratory in Canada, at 6010 meters of water equivalent depth. Here we report results from that run.

\begin{figure}
\includegraphics[width=250 pt,trim=0 0 0 0,clip=true]{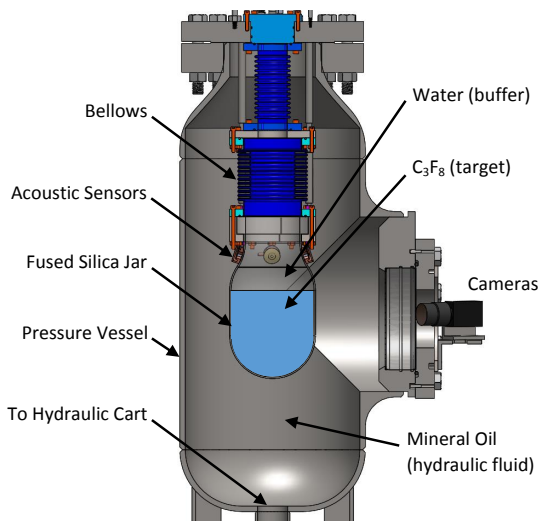}
\caption{\label{fig:schematic} A schematic of the PICO-2L bubble chamber. 
}
\end{figure}

The bubble chamber (called PICO-2L) deployed in this experiment was very similar to the 2 liter chambers described previously~\cite{previousPRL,PRD}, primarily consisting of a fused-silica jar sealed to a flexible, stainless steel bellows, all immersed in a pressure vessel filled with hydraulic fluid. The jar was filled with $2.90\pm0.01$ kg of C$_3$F$_8$  (2.09 liters of fluid at a density of 1.39 kg/L at 12$^{\circ}$C and 30~psia), as measured by a scale, with the uncertainty due to losses in the fill lines and electronic noise in the scale readout. To isolate it from contact with any stainless steel surfaces or seals, the C$_3$F$_8$ is topped with a water buffer layer. A schematic of the chamber is shown in Fig.~\ref{fig:schematic}. 

  Three 
  lead zirconate (PZT) piezoelectric acoustic transducers epoxied to the exterior of the fused-silica jar recorded the acoustic emissions from bubble nucleations~\cite{GlaserPlink}.  Previously, high levels of radioactivity in the transducers provided a measurable neutron rate~\cite{PRD}. For PICO-2L, we developed PZT sensors from source material with a factor 100 reduction in radioactivity.  The acoustic signals were digitized with a sampling rate of $2.5$~MHz. Two VGA resolution CCD cameras photographed the chamber at a rate of $100$ frames per second.

 \begin{table*}[ht] 
 \begin{center} 
 \begin{tabular}{|c|c|c|c|c|c|} \hline
T ($^{\circ}$C) & P (psia) & Seitz threshold, $E_T$ (keV) & Livetime (d) & WIMP exposure (kg-d) & No. of candidate events\\\hline
    14.2 & 31.1 & $3.2 \pm 0.2(\mathrm{exp}) \pm 0.2(\mathrm{th})$ &32.2 & 74.8 & 9 \\
  12.2 & 31.1 & $4.4 \pm 0.3(\mathrm{exp}) \pm 0.3(\mathrm{th})$&7.5 & 16.8 &  0 \\
  11.6 & 36.1 & $6.1 \pm 0.3(\mathrm{exp}) \pm 0.3(\mathrm{th})$&39.7 & 82.2 &  3 \\
  11.6 & 41.1 & $8.1 \pm 0.5(\mathrm{exp}) \pm 0.4(\mathrm{th})$&18.2 & 37.8 & 0\\
\hline\hline
\end{tabular}
 \caption{\label{table:opcond}Table describing the four operating conditions and their associated exposures. The experimental uncertainty on the threshold comes from uncertainties on the temperature (0.3$^{\circ}$C) and pressure (0.7~psi), while the theoretical uncertainty comes from the thermodynamic properties of C$_3$F$_8$ (primarily the surface tension).}
 \end{center}
\end{table*}

The PICO-2L event cycle was similar to that described previously~\cite{PRD}. The chamber was operated at four pressure and temperature combinations, listed in Table~\ref{table:opcond}. The pressure and temperature determine the conditions for radiation-induced bubble nucleation, approximated by Seitz's ``hot spike'' model~\cite{seitztheory} in which the particle interaction must provide the energy necessary to produce a critically-sized bubble. We follow the method described in~\cite{CIRTE} to calculate the Seitz threshold for bubble nucleation ($E_T$) for each run condition of PICO-2L and for the remainder of the letter refer to each data set by the calculated threshold. We quote both experimental and theoretical uncertainties in $E_T$, the former from uncertainties in the pressure and temperature of the target fluid, and the latter from uncertainties in the surface tension for very small bubbles~\cite{CIRTE}. 

The chamber was exposed to an AmBe calibration source ten times during the run to monitor the detector response to nuclear recoils. All calibration data were handscanned to check bubble multiplicities, and hand-scanned single bubble events were used to determine the data cleaning cut efficiencies.

    The data analysis begins with an image reconstruction algorithm to identify bubbles and their locations in 3D space. An optical-based fiducial volume cut is derived from neutron calibration data such that $1\%$ or fewer of wall or surface events, defined as events located on the glass jar or at the interface between the C$_3$F$_8$ and water buffer respectively, are reconstructed as bulk events, defined as bubbles that do not touch either the glass or water. The efficiency of the optical fiducial cut is determined to be $0.82\pm0.01$ by volume (all error bars on cut efficiencies are $1\sigma$ and represent total uncertainties).

 In~\cite{PRD}, the rate-of-pressure-rise during an event was used as a highly efficient fiducial volume cut, as bubble growth is affected by proximity to the jar or the liquid interface. A similar analysis was implemented in PICO-2L with an efficiency of $0.92\pm0.02$, in agreement with~\cite{PRD}.  The pressure-rise analysis could not be applied to all data as improvements to the PICO-2L data acquisition system and hydraulic cart reduced the time between trigger and compression, stopping bubble growth before the pressure could increase significantly. A trigger delay of 10-40 ms was imposed for most of the low threshold data to allow more time for the bubble to evolve, enabling use of the pressure rise cut. For the higher threshold data without the trigger delay, the optical fiducial cut is used.

\begin{figure}[ht] 
\includegraphics[width=250 pt]
{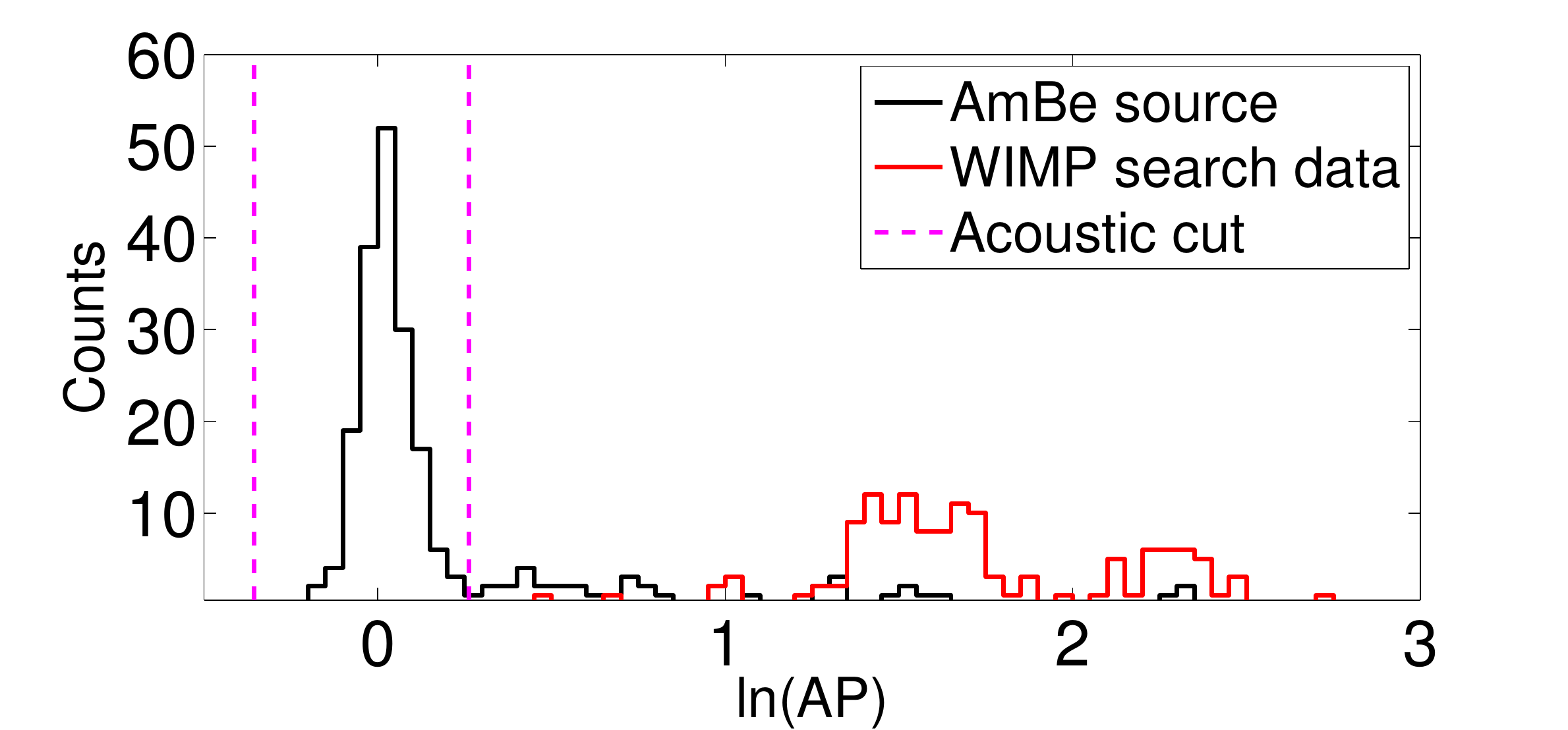}
\caption{\label{fig:APdistribution} 
$AP$ distributions for neutron calibration data (black) and WIMP search data (red) at 4.4 keV threshold. Note that the x-axis shows $\ln(AP)$. As discussed in the text, alphas from the $^{222}$Rn decay chain can be identified by their time signature and populate the two peaks in the WIMP search data at high $AP$, with higher energy alphas from $^{214}$Po producing larger acoustic signals.
}
\end{figure}

The acoustic analysis follows the procedure described in~\cite{PRD} to define  $AP$, a measurement of the acoustic power released in an event. 
Figure~\ref{fig:APdistribution} shows the $AP$ distributions for calibration and WIMP search data at a threshold of 4.4 keV. The $AP$ distribution is normalized to have a value of unity at the nuclear recoil peak observed in the AmBe data, and an acoustic cut is applied to select these events.  For the two low threshold data sets, we adopt the same acoustic cut as in~\cite{previousPRL,PRD}, such that $0.7 < AP < 1.3$. Because of the decreased acoustic signal at higher operating pressure, the width of the calibration peak at 6.1 keV threshold is a factor of 1.5 larger than at low thresholds; the acceptance region for this data set is chosen such that the difference between the cut value and the mean divided by the resolution is the same as for low thresholds ($0.55 < AP < 1.45$). At 8.1 keV threshold, some neutron-induced events are too quiet to be registered acoustically, so all events with $AP < 2$ are counted as nuclear recoil events. The acceptance of these cuts for neutron-induced single bubble events was statistically indistinguishable for all data sets with a value of $0.91 \pm 0.01$.

 A set of quality cuts is applied to all data to eliminate events with excessive acoustic noise, events where the cameras failed to capture the initiation of the bubble, and events in which the optical reconstruction algorithm failed to converge. The total efficiency of the data quality cuts is $0.961 \pm0.003$.  The total acceptance for neutron-induced, single nuclear recoils including fiducial, acoustic and data quality cuts is  $0.80 \pm 0.02$ for data with the trigger delay and the pressure-rise based fiducial cut, decreasing to $0.72 \pm 0.02$ for the optical fiducial cut.

One of the main strengths of the superheated fluid detectors is their insensitivity to electronic recoils. The PICO-2L chamber was exposed to a 1~mCi $^{133}$Ba source to confirm this behavior in C$_3$F$_8$. With no candidate events observed during the gamma exposure at 3.2 keV, the probability for a gamma interaction to nucleate a bubble was determined to be less than $3.5\times10^{-10}$ at 90\% C.L. by performing a Geant4~\cite{GEANT4} Monte Carlo simulation of the source and counting the total number of above-threshold interactions of any kind in the active target. Combining these results with a dedicated NaI measurement of the gamma flux at the location of the chamber in the absence of any sources ~\cite{DrewThesis}, we expect electronic recoils to produce fewer than 0.05 events in the PICO-2L WIMP search data.

A second key method for background rejection in superheated detectors is the acoustic rejection of alpha decays~\cite{PICASSOdiscrimination,previousPRL,PRD,simple2014}. PICO-2L observed a rate of high-$AP$ events at 4.4 keV threshold immediately after the initial fill that decayed with a half-life consistent with that of $^{222}$Rn to a steady state of about 4 events/day.  None of the high acoustic power events leak into the nuclear recoil acceptance band in that data set, confirming that acoustic alpha rejection is present in the C$_3$F$_8$ target. The 4.4 keV data provide a statistics-limited, $90\%$ lower limit on the alpha rejection in PICO-2L of $98.2\%$. 

In addition to the acoustic discrimination, PICO-2L data show a dependence of $AP$ on alpha energy that was not previously observed in CF$_3$I. At low threshold, two distinct peaks appear at high $AP$ (see Fig.~\ref{fig:APdistribution}).  The time structure of the high-$AP$ peaks is consistent with that of the fast radon chain ($^{222}$Rn, $^{218}$Po, and $^{214}$Po decays having energies of 5.5 MeV, 6.0 MeV, and 7.7 MeV, respectively). The events in the louder peak come primarily from the third event in the chain, the high energy $^{214}$Po decay. To our knowledge, this constitutes a first instance of particle energy spectroscopy using acoustic methods.

Background neutrons produced primarily by ($\alpha$,n) and spontaneous fission from nearby $^{238}$U and $^{232}$Th can produce both single and multiple bubble events. 
We perform a detailed Monte Carlo simulation of the detector to model the neutron backgrounds, predicting 0.9(1.6) single(multiple) bubble events in the entire data set, for an event rate of 0.004(0.006) cts/kg/day, with a total uncertainty of $50\%$. There were no multiple bubble events observed in the WIMP search data, providing a $90\%$ C.L. upper limit of 0.008 cts/kg/day, consistent with the background model.

The sensitivity of the experiment to dark matter depends crucially on the efficiency with which nuclear recoils at a given energy produce bubbles. The classical Seitz model~\cite{seitztheory} indicates that nuclear recoils of energy greater than $E_T$ will create bubbles with 100\% efficiency, but past results show that the model does not accurately describe the efficiency for detecting low energy carbon and fluorine recoils in CF$_3$I~\cite{PRD,AlanIDM}.
This breakdown is attributed to the relatively large size of carbon and fluorine recoil tracks in CF$_3$I, as bubble nucleation only occurs if the energy deposition is contained within a critical bubble size. Iodine recoils in CF$_3$I have much shorter tracks and have been shown to more closely match the Seitz model predictions~\cite{CIRTE}. Simulations of nuclear recoil track geometries using the Stopping Range of Ions in Matter (SRIM) package~\cite{TRIM3} as well as measurements in C$_4$F$_{10}$~\cite{picassoCal} indicate that fluorine recoils in C$_3$F$_8$ are also in the regime where the Seitz model is a close approximation for bubble nucleation.

To confirm this expectation, we performed neutron calibrations \emph{in situ} in the PICO-2L chamber with an AmBe neutron source. We also deployed a $\sim$30-ml C$_3$F$_8$ bubble chamber at the Tandem Van de Graaff facility at the University of Montreal, using well-defined resonances in the $^{51}$V(p,n)$^{51}$Cr reaction to produce monoenergetic 61- and 97-keV neutrons.  Each of the three neutron calibration experiments is simulated in MCNP~\cite{MCNP-PoliMi} using updated differential cross sections for elastic scattering on fluorine~\cite{AlansPaper}.

\begin{figure}[ht!]
\begin{center}
\includegraphics[width=0.5\textwidth, trim = 0 0 0 0, clip = true]{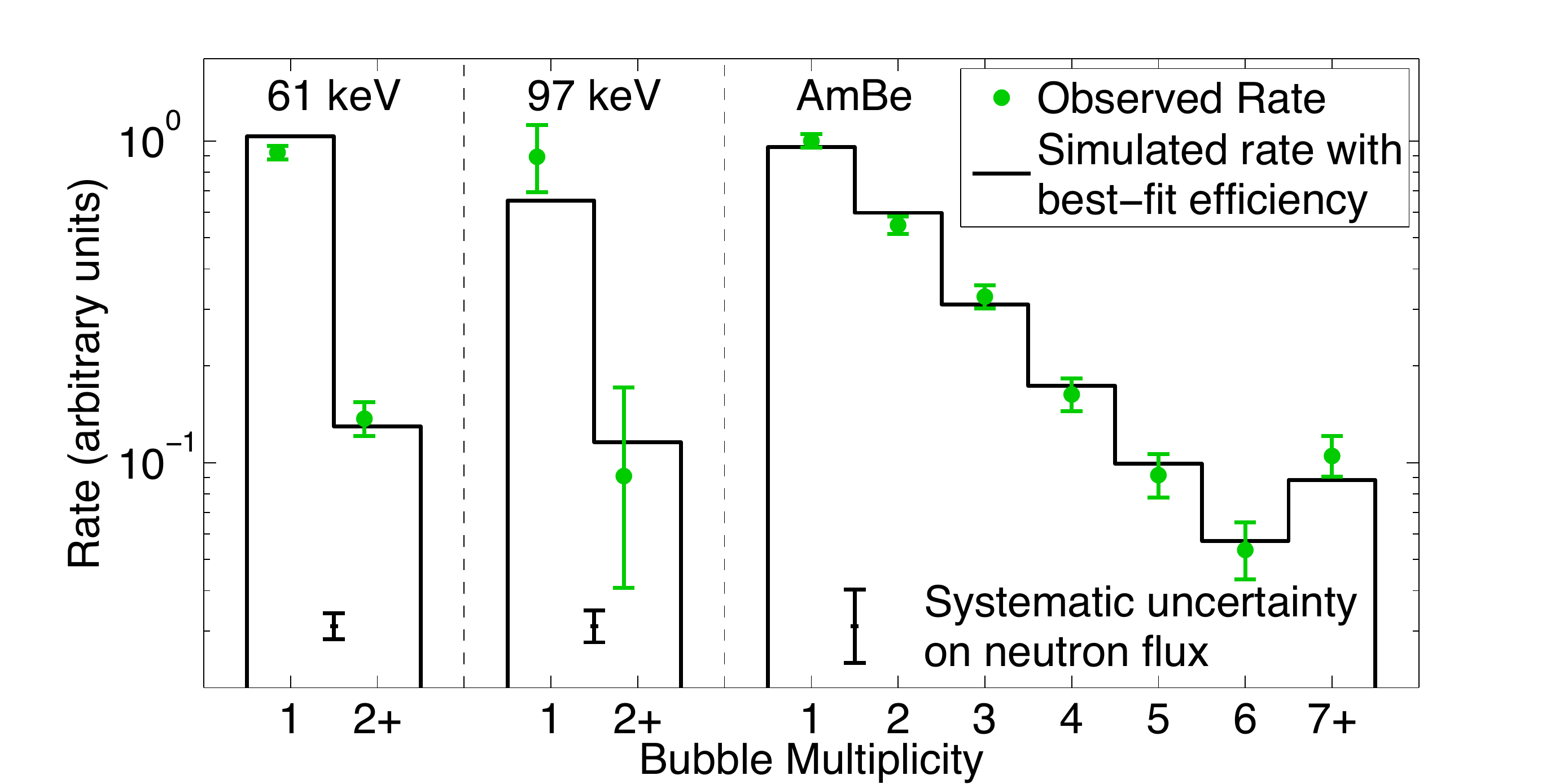}
\caption{\label{fig:NeutronMC} The green points show the observed rates of single and multiple bubbles for the calibration sources at a thermodynamic threshold of 3.2~keV. Green error bars indicate statistical uncertainties, and the black error bars at the bottom show the systematic uncertainty on the neutron flux (a flat percent uncertainty that is common to all multiplicities in a given data set at the $10\%$, $12\%$, and $30\%$ level for 61 keV, 97 keV and AmBe data, respectively).  The black histograms show the predicted rates from the simulation given the best-fit efficiency model derived from all calibration data. 
}
\end{center}
\end{figure}

\begin{figure}[ht!]
\begin{center}
\includegraphics[width=0.45\textwidth, trim = 0 0 0 0, clip = true]{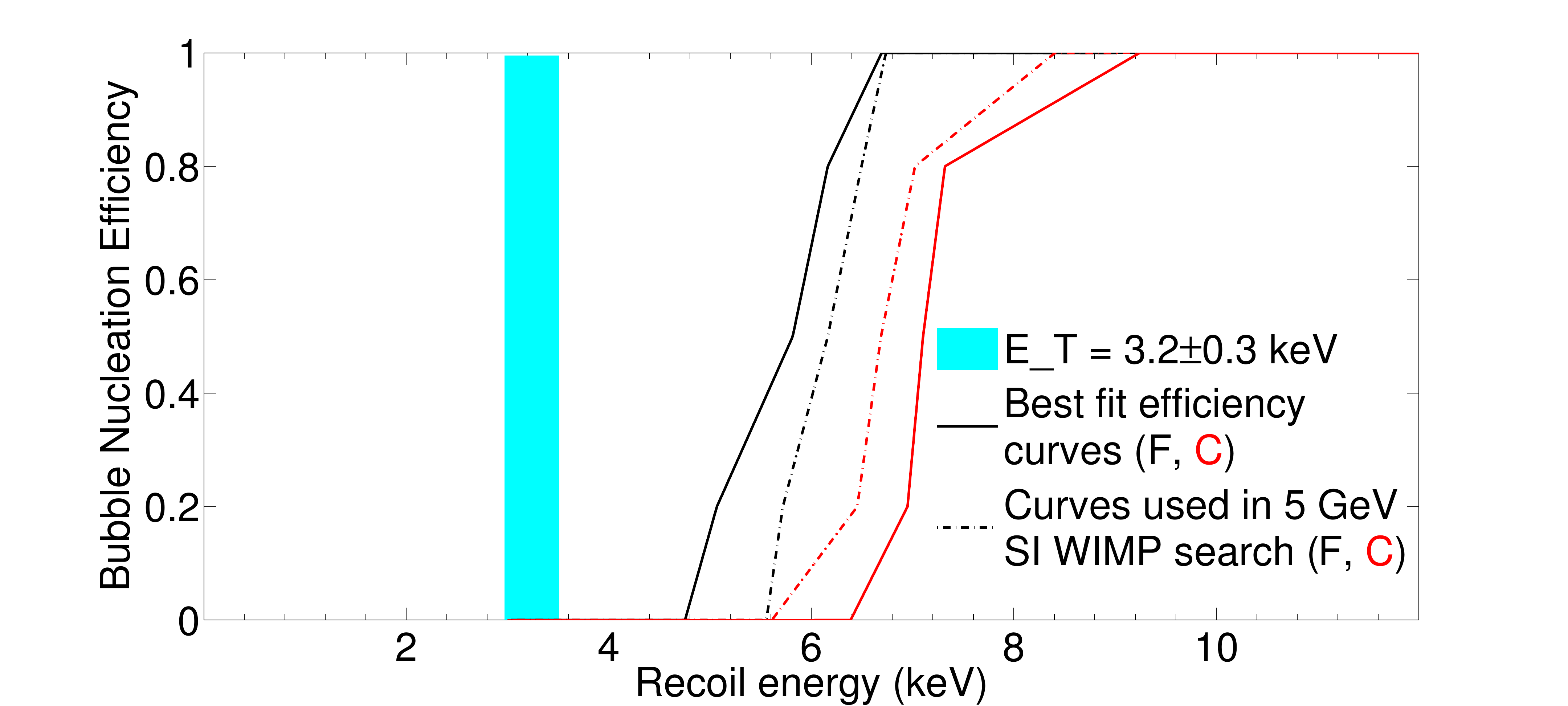}
\caption{\label{fig:efficiency} The best-fit fluorine (black) and carbon (red) efficiency curves for 3.2 keV data are shown by the solid lines. The dashed lines show the curves used to determine sensitivity for a 5 GeV SI WIMP. The light blue band shows the calculated Seitz threshold with the experimental and theoretical uncertainties from Table~\ref{table:opcond} added in quadrature. }
\end{center}
\end{figure}

A single calibration point, i.e., a bubble rate measured at a given thermodynamic threshold and produced by a single spectrum of nuclear recoil energies, can in general be fit by a family of possible nucleation efficiency curves.  
In this analysis, the fluorine and carbon efficiency curves at each threshold are fit by monotonically increasing, piecewise linear functions to allow for a variety of different efficiency shapes, with no reference to the Seitz theory except that bubble nucleation cannot occur for recoil energies below $E_T$ (subject to the experimental uncertainties). In addition, the carbon efficiency is assumed to be less than or equal to the fluorine efficiency at a given recoil energy from track geometry considerations. Figure~\ref{fig:NeutronMC} shows the observed rates of single and multiple bubbles for the AmBe and test beam sources compared to the best-fit efficiency model at a thermodynamic threshold of 3.2 keV.  The best-fit efficiency curves for fluorine and carbon at 3.2 keV are shown by the solid lines in Fig.~\ref{fig:efficiency}.

We take a conservative approach when determining the sensitivity of PICO-2L to dark matter.  For each WIMP mass and coupling, we select the pair of fluorine and carbon efficiency curves giving the worst sensitivity for that particular WIMP that is consistent with the calibrations at $1\sigma$. As an example, the dashed lines in Fig.~\ref{fig:efficiency} show the actual efficiency curves used to determine the sensitivity of the experiment for a 5 GeV SI WIMP for the $E_T = 3.2$ keV data set. For this case, where most of the sensitivity to WIMPs comes from the lowest energy fluorine recoils, our conservative approach uses a weaker response to fluorine relative to the best-fit case (e.g. the turn-on is shifted to slightly higher energies). Because the total rate in the calibration data is unchanged, the fit compensates for the weaker fluorine response by assuming a larger contribution from carbon. The difference between the solid and dashed lines is small, attesting to how well the calibration data constrain the C$_3$F$_8$ response. 

As shown in Table~\ref{table:opcond}, WIMP search data were taken at four different thresholds, with most data coming at thresholds of 3.2 keV and 6.1 keV. There are nine candidate events within the $AP$ acceptance region at 3.2 keV and three candidate events at 6.1 keV, with no candidate events observed at 4.4 and 8.1 keV. All 12 candidate events were hand scanned and found to be well reconstructed, bulk events.

\begin{figure}[ht!]
\begin{center}
\includegraphics[width=0.5\textwidth, trim = 0 0 0 0, clip = true]{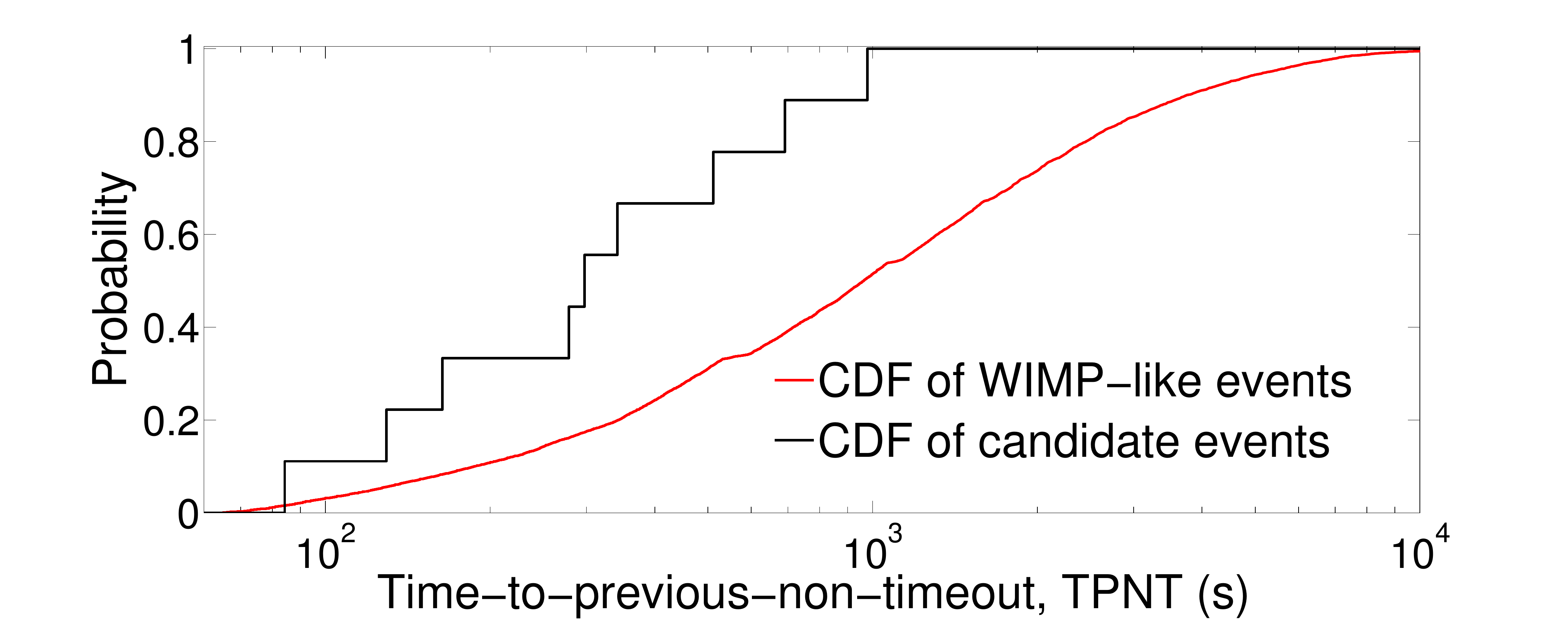}
\caption{\label{fig:timing} The CDF of the time to previous non-timeout (TPNT) for events with random timing (simulated WIMP-like events) and the 3.2 keV candidate events. The two distributions are not consistent with each other. }
\end{center}
\end{figure}

In~\cite{PRD}, WIMP-candidate events were observed exhibiting correlations with events in previous expansions, and the candidate events in PICO-2L exhibit similar correlations. To explore this anomaly further, simulated events with random timing are populated into the actual data to model the expected timing distribution of a potential WIMP signal.  Figure~\ref{fig:timing} shows the cumulative distribution function (CDF) of the time to previous non-timeout (TPNT) for a randomly distributed sample, along with the TPNT for each candidate event at 3.2 keV. A Kolmogorov-Smirnov test comparing the two samples returns a p-value of 0.04 that they are drawn from the same distribution.  Given these results, the candidate events are not treated as evidence for a dark matter signal but instead as an unknown background. Studies are now underway to test hypotheses for the source of these events.

The correlation of the candidate events with previous bubbles can be used to set a stronger constraint on WIMP-nucleon scattering by applying a cut on TPNT. Since there is no valid basis for setting the cut value {\it a priori}, a method based closely on the optimum interval method~\cite{optimuminterval} is used to provide a true upper limit with TPNT cuts for each WIMP mass optimized simultaneously over all four operating thresholds. The optimum cuts remove all 12 candidate events at each WIMP mass, while retaining 49--63\% of the efficiency weighted exposure, with the range due to changes in the relative weighting of the four threshold conditions for different WIMP masses. If the optimum cuts had simply been set {\it a posteriori}, without applying the tuning penalty inherent in the optimization method, the cross section limits would be a factor of 1.2--2.4 lower than reported here, with the bigger factor applying to higher WIMP masses.

\begin{figure}[htbp!]

\includegraphics[width=250 pt]{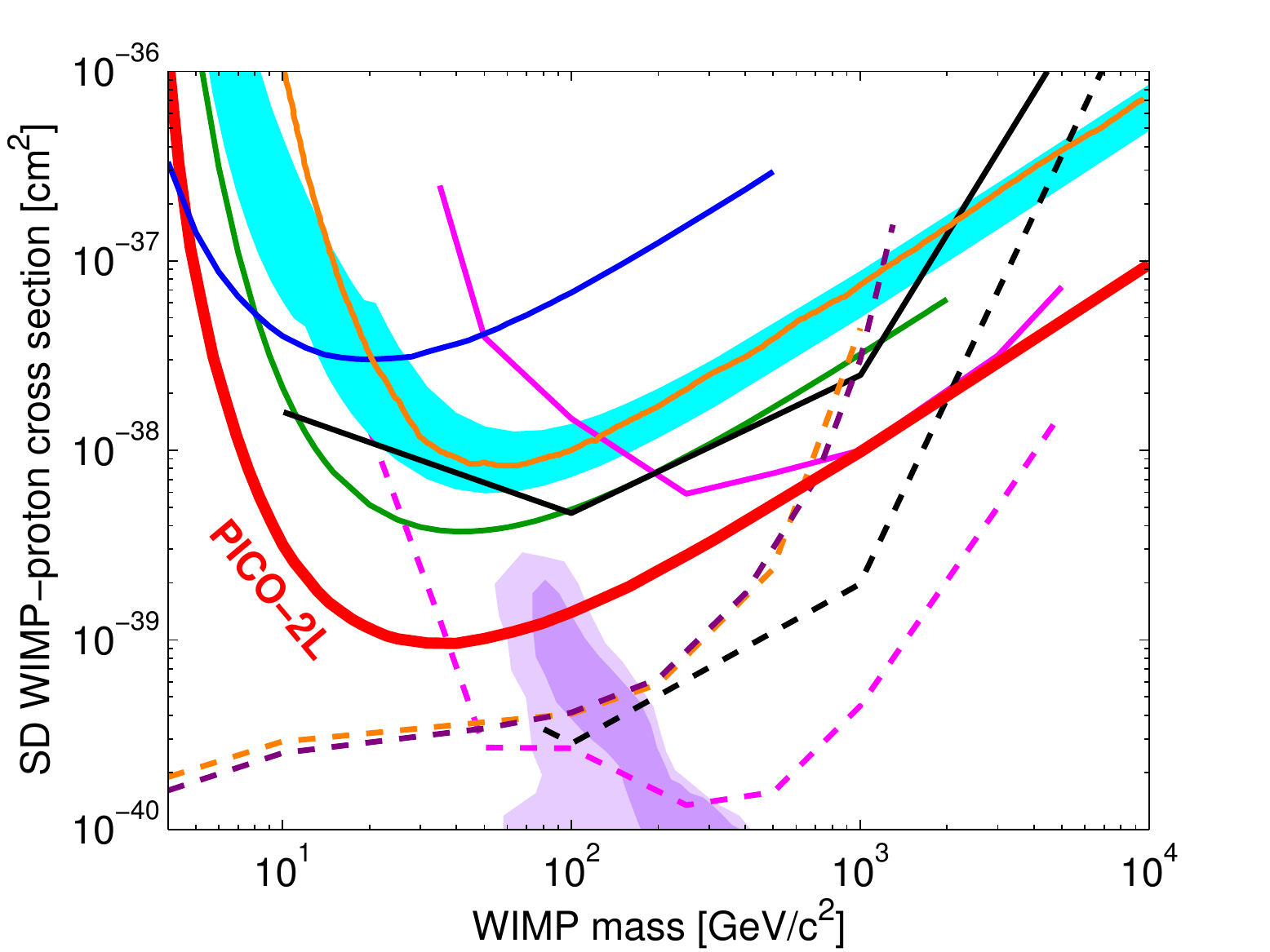}
\caption{\label{Spin_Dependent_Limit}
The $90\%$ C.L. limit on the SD WIMP-proton cross section from PICO-2L is plotted in red, along with limits from COUPP (light blue region), PICASSO (dark blue), SIMPLE (green), XENON100 (orange), IceCube (dashed and solid pink), SuperK (dashed and solid black), CMS (dashed orange), and ATLAS (dashed purple)~\cite{PRD,PICASSOlimit,simple2014,XENON100_SD,ICECUBElimit,SKlimit,CMSlimit, ATLASheavyquark}. For the IceCube and SuperK results, the dashed lines assume annihilation to $W$-pairs while the solid lines assume annihilation to $b$-quarks. Comparable limits assuming these and other annihilation channels are set by the ANTARES, Baikal and Baksan neutrino telescopes~\cite{Antares,Baksan,Baikal}. The CMS and ATLAS limits assume an effective field theory, valid for a heavy mediator. The purple region represents parameter space of the CMSSM model of~\cite{SDblob}. 
}
\end{figure}

\begin{figure}[htbp!]
\includegraphics[width=250 pt]{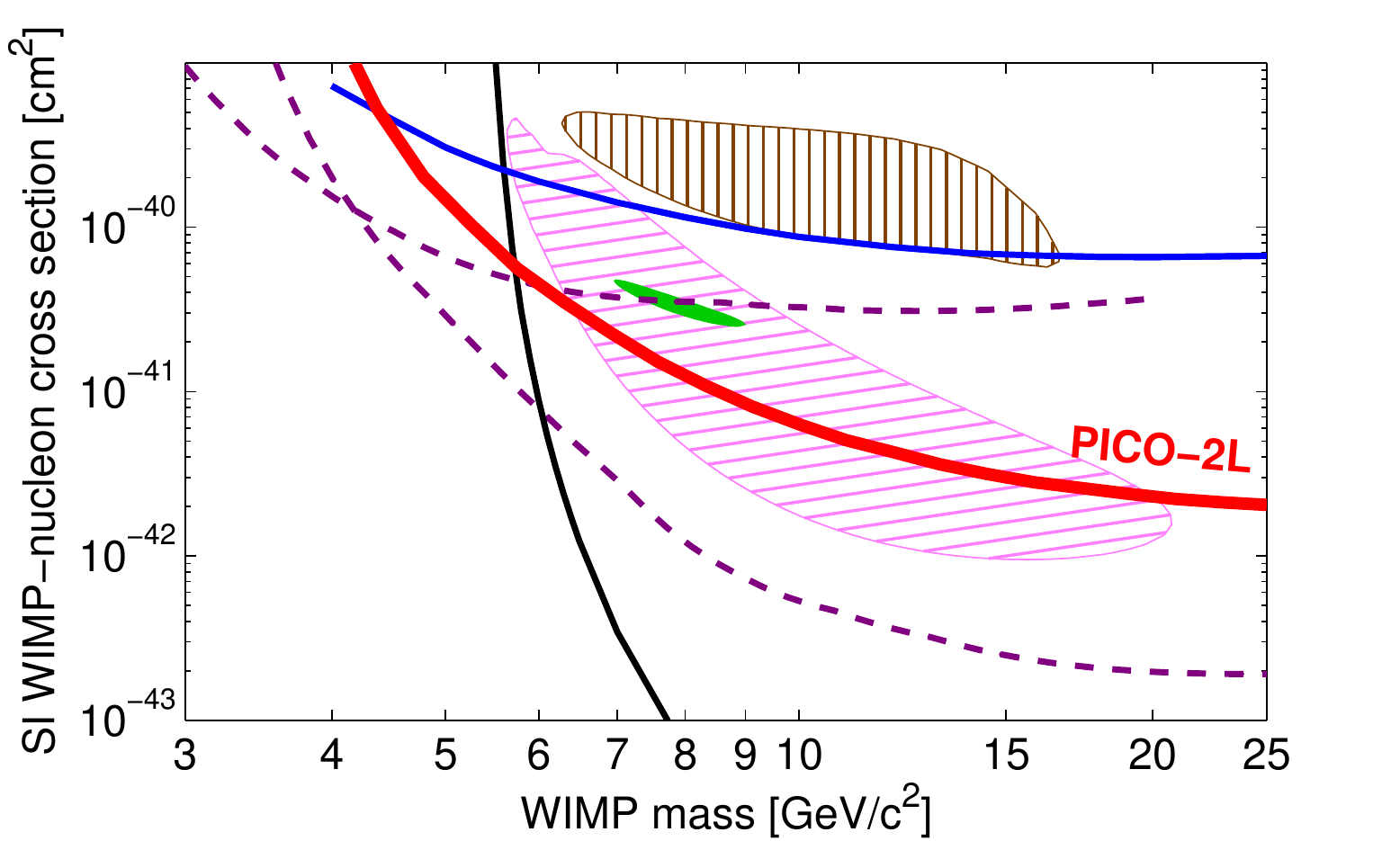}
\caption{\label{Spin_INdependent_Limit} 
The $90\%$ C.L. limit on the SI WIMP-nucleon cross section from PICO-2L is plotted in red, along with limits from PICASSO (blue), LUX (black), CDMS-lite and SuperCDMS (dashed purple)~\cite{PICASSOlimit, LUX,CDMSlite,SuperCDMS}. Similar limits that are not shown for clarity are set by XENON10, XENON100 and CRESST-II~\cite{XENON10,XENON100,CRESSTII}. Allowed regions from DAMA (hashed brown), CoGeNT (solid green), and CDMS-Si (hashed pink) are also shown~\cite{DAMA, CoGeNT, CDMSSi}. 
}
\end{figure}

The limit calculations assume the standard halo parametrization~\cite{lewinandsmith}, with $\rho_D = 0.3$ GeV $c^{-2}$ cm$^{-3}$, $v_\mathrm{esc}=544$ km/s, $v_\mathrm{Earth}=232$ km/s, $v_0=220$ km/s, and the spin-dependent parameters from~\cite{spindependentcouplings}, and the resulting $90\%$ C.L. limit plots for spin-independent WIMP-nucleon and spin-dependent WIMP-proton cross sections are presented in Figs.~\ref{Spin_INdependent_Limit} and~\ref{Spin_Dependent_Limit}. Using the same parameters as in~\cite{lewinandsmith} would yield approximately $5-20\%$ stronger limits depending on the WIMP mass. The results shown here represent the most stringent constraint on SD WIMP-proton scattering from a direct detection experiment and the first time supersymmetric parameter space has been probed by direct detection in the SD-proton channel (e.g. the purple region, taken from~\cite{SDblob}).

\begin{acknowledgments}

The PICO Collaboration would like to thank SNOLAB and its staff for providing an exceptional underground laboratory space and invaluable technical support.  This material is based upon work supported by the U.S. Department of
Energy, Office of Science, Office of High Energy Physics under award DE-SC-0012161. Fermi National
Accelerator Laboratory is operated by Fermi Research Alliance, LLC under
Contract No. De-AC02-07CH11359.  Part of the research described in this
paper was conducted under the Ultra Sensitive Nuclear Measurements
Initiative at Pacific Northwest National Laboratory, a multiprogram
national laboratory operated by Battelle for the U.S. Department of
Energy.  

We acknowledge
the National Science Foundation for their support including Grants No. PHY-1242637, No. PHY-0919526, and No. PHY-1205987. We acknowledge the support of the National Sciences and Engineering Research Council of Canada (NSERC) and the Canada Foundation for Innovation (CFI). We also acknowledge support from the Kavli Institute for Cosmological Physics at the University of Chicago. We acknowledge the financial support of the Spanish Ministerio de Econom\'ia y Competitividad, Consolider MultiDark CSD2009-00064 Grant. We acknowledge support from the Department of Atomic Energy (DAE), Government of India, under the Center for AstroParticle Physics-II project at SINP, Kolkata. We acknowledge the Czech Ministry of Education, Youth and Sports, Grant No. LM2011027.  We acknowledge technical assistance from Fermilab's Computing, Particle Physics, and Accelerator Divisions, and from A. Behnke at IUSB. 

\end{acknowledgments}

\end{document}